\documentclass[nofootinbib,floats,floatfix,eqsecnum,prd,aps]{revtex4}
\usepackage{dcolumn,epsfig,minitoc}
\usepackage{amssymb,amsmath}
\usepackage[usenames]{color}

\def\a{\alpha}
\def\b{\beta}

\def\l{\lambda}

\def\l{\lambda}

\def\o1{$O_{1}$}
\def\o2{$O_{2}$}

\newcommand{\beq}{\begin{equation}}
\newcommand{\eeq}{\end{equation}}
\newcommand{\bea}{\begin{eqnarray}}
\newcommand{\eea}{\end{eqnarray}}

\begin{document}

\title{Hyperscaling violation for scalar black branes in arbitrary dimensions%
}
\author{Mariano Cadoni, Matteo Serra}
\affiliation{Dipartimento di Fisica, Universit\`a di Cagliari and INFN, Sezione di
Cagliari - Cittadella Universitaria, 09042 Monserrato, Italy. }
\date{\today}

\begin{abstract}
We extend to black branes (BB) in arbitrary dimensions the results of Ref. 
\cite{Cadoni:2012uf} about hyperscaling violation and phase transition for scalar black $%
2$-branes. We derive the analytic form of the $(d+1)$-dimensional scalar
soliton interpolating between a conformal invariant AdS$_{d+2}$ vacuum in
the infrared  and a scale covariant metric in the ultraviolet. 
We show that the
thermodynamical system undergoes a phase transition between Schwarzschild-AdS%
$_{d+2}$ and a scalar-dressed BB. We calculate the critical exponent $z$ and
the hyperscaling violation parameter $\theta $ in the two phases. We show
that our scalar BB solutions generically emerge as compactifications of $p-$%
brane solutions of supergravity theories. We also derive the short distance
form of the correlators for the  scalar operators corresponding to  an UV exponential
potential supporting our 
black brane solution. We show that also for negative $%
\theta $ these correlators have  a short distance power-law behavior.
\end{abstract}

\maketitle






\section{Introduction}

Recent investigations on the application of the anti-de Sitter/conformal
field theory (AdS/CFT) correspondence to strongly interacting quantum field
theories (QFT) have emphasized the importance played by non-AdS
gravitational backgrounds and the related dual nonconformal QFTs \cite%
{Charmousis:2009xr, Cadoni:2009xm,Goldstein:2009cv,
Goldstein:2010aw,
Charmousis:2010zz,Gouteraux:2011ce,
Cadoni:2011kv,Lee:2011zzf,Iizuka:2011hg,Ammon:2012je,Kim:2012nb,
Dong:2012se, Wu:2012fk}.

The standard setup for this kind of holographic applications is a black
brane in a AdS background endowed with non trivial scalar field
configurations and finite electromagnetic charge density. It has been shown
that this produces  a rich phenomenology in the dual QFT, such as
spontaneous  symmetry breaking, phase transitions and non-trivial transport properties 
\cite{Hartnoll:2008vx,Hartnoll:2008kx,
Horowitz:2008bn,Herzog:2009xv,Hartnoll:2009sz,Charmousis:2009xr,
Cadoni:2009xm,Goldstein:2009cv,Goldstein:2010aw,
Charmousis:2010zz,Bertoldi:2010ca, Gouteraux:2011ce,Iizuka:2011hg,
Cadoni:2011kv,Lee:2011zzf,Kim:2012nb,
Dong:2012se, 
Wu:2012fk,Hashimoto:2012ti,Ammon:2012je,Iizuka:2012wt,Dey:2012rs,
Park:2012cu,Myung:2012cb,Mignemi:2012rd}.

In the case of nonminimal, exponential, coupling between the scalar field
and the Maxwell tensor  the bulk gravity allows for 
extremal, near-horizon,  solutions which
break the  conformal symmetry of the AdS vacuum \cite%
{Goldstein:2009cv,Cadoni:2009xm,Goldstein:2010aw,Charmousis:2010zz,
Cadoni:2011kv,Bertoldi:2011zr,Iizuka:2011hg,Kim:2012nb,Dong:2012se,
Wu:2012fk}. It has been realized that  these IR metrics belong to a
general class of metrics that are not scale invariant but only scale
covariant \cite{Gouteraux:2011ce,Cadoni:2011nq,Cadoni:2011kv} and lead to
hyperscaling violation in the dual field theory \cite%
{Huijse:2011ef,Dong:2012se,Kim:2012nb,Hartnoll:2012wm,Dey:2012fi,
Sadeghi:2012vv,Bhattacharya:2012zu,Alishahiha:2012cm,Alishahiha:2012qu}. They are characterized by two
parameters, the critical exponent $z$ and the hyperscaling violation
parameter $\theta$, which characterize both the transformation weight of
the infinitesimal length $ds$ under scale transformations and the scaling
behavior of the free energy as function of the temperature \cite%
{Huijse:2011ef,Dong:2012se}.

Scale covariant metrics are a very promising framework for the holographic
description of quantum phase transitions and hyperscaling violation in
condensed matter critical systems (e.g Ising models) \cite{Fisher:1986zz}.

An important holographic feature of scale-covariant metrics is the emergence
of a length-scale in the IR \cite{Dong:2012se}, which decouples in the UV
when the theory has an UV fixed point. The emergence of a length-scale in
the IR is crucial for the description of Fermi surfaces and for the related
area-law violation of entanglement entropy \cite
{Dong:2012se,Huijse:2011ef,Shaghoulian:2011aa,Ogawa:2011bz,Li:2012uua,
Kulaxizi:2012gy}.

The standard framework for obtaining, dynamically, scale -covariant metrics
in the IR, is given by Einstein-scalar gravity, possibly coupled --
minimally or non-minimally -- to a $U(1)$ field. The self-interaction
potential $V(\phi )$ for the scalar field $\phi $ must have a negative local
maximum at $\phi =0$, with a corresponding scalar tachyonic excitation whose
mass is slightly above the Breitenlohner-Freedman (BF) bound. Under suitable
conditions, usually an exponential behavior of the potential and/or
scalar-Maxwell tensor coupling functions, the theory admits black brane
solutions with scalar hair that in the near-extremal regime approach the
scale covariant metrics.

In a recent paper \cite{Cadoni:2012uf} it has been shown that this framework is not the
only possible way to produce scale covariant BB. These solutions can be also
obtained from Einstein-scalar gravity with a positive squared  mass 
for the scalar,
when the potential behaves exponentially in the asymptotic region of the
spacetime \cite{Cadoni:2011nq,Cadoni:2011yj}. Standard no-hair theorems for Einstein-scalar gravity
model with positive squared mass for the scalar field 
\cite{Townsend:1984iu,Torii:2001pg,Hertog:2006rr} can be circumvented by
giving up the condition that the BB solution has AdS asymptotics \cite
{Cadoni:2011nq,Cadoni:2011yj}.

Although BB solutions with scale-covariant asymptotics have been explicitly
derived for particular four-dimensional (4D) Einsten-scalar gravity models, their existence is
a rather generic feature of a broad class of 4D models \cite{Cadoni:2012uf}. Moreover,
in the extremal limit the BB solution reduces to a fully regular scalar
soliton, which interpolates between an AdS$_{4}$ vacuum in the near-horizon
region and a scale covariant solution in the asymptotic region.

These results allow to realize an alternative scenario, which 
exchanges  IR and UV regions. In the dual QFT we have an infrared fixed point, 
corresponding to the AdS vacuum, whereas in the UV  regime we have
hyperscaling violation. 

Detailed investigation of the symmetries and thermodynamics of these BB
solutions revealed rather interesting and intriguing features. The
thermodynamical phase diagram of the system is characterized by the presence
of different phases. Above a critical temperature $T_{c}$ the scalar dressed
BB brane becomes energetically preferred with respect to the
Schwarzschild-AdS$_{4}$ (SAdS) solution and the thermodynamical system
undergoes a first order phase transition. Moreover, for some values of the
parameters characterizing the model, at low temperatures different phases may
coexist. In the dual QFT the scalar dressed, stable, BB corresponds to a phase with a
negative hyperscaling violation parameter $\theta $. Although negative
values of $\theta $ do not have analogous in condensed matter system, they
are consistent with the null energy conditions for the bulk stress-energy
tensor. Moreover they also arise in string theory and supergravity
constructions 
\cite{Perlmutter:2010qu,Narayan:2012hk,Dong:2012se,Dey:2012rs,Perlmutter:2012he}.

The purpose of this paper is the generalization of the results of Ref. \cite
{Cadoni:2012uf} concerning $2$-branes to branes of arbitrary dimensions. We will show
that basically all the results of Ref. \cite{Cadoni:2012uf} can be generalized to
arbitrary dimensions and therefore generically hold for $d$-branes. Apart
from generalizing the results of Ref. \cite{Cadoni:2012uf} to arbitrary dimensions, we
will also show that our scalar BB solutions can be obtained in several ways
as compactifications of $p$-brane solutions of supergravity (SUGRA) theories.

Finally, we will be concerned with some holographic features of QFTs with
negative hyperscaling violation parameter $\theta$. Extending the results of
Ref. \cite{Dong:2012se}, which hold for positive $\theta$ and for a 
massive scalar field, we will derive the short
distance form of the correlators for scalar operators 
corresponding to an UV exponential
potential supporting our 
black brane solution. We show that for negative $%
\theta $ these correlators have  a short distance power-law behavior.

The paper is organized as follows. In Sect. \ref{sect:AA} we present our
Einstein-scalar gravity model, derive the BB solutions with scale-covariant
asymptotics and discuss their solitonic extremal limit. In Sect. \ref%
{sect:AB} we show how our BB solutions can be obtained as compactifications
of $p$-brane solutions of SUGRA theories. The thermodynamics of our
solutions is investigated in Sect. \ref{sect:AC}. The symmetries of the BB
are discussed in Sect. \ref{sect:AD}, where also critical exponent and
hyperscaling violation parameters are calculated. In Sect. \ref{sect:AEE} we
extend our investigation to general models whose potential behaves
exponentially in the asymptotic region. In Sect. \ref{sect:AE} we study
holographic properties of our BB solution and in particular the two-point
function of scalar operators in the dual QFT. Finally, in Sect. \ref{sect:AF}
we state our conclusions.

\section{Einstein-scalar gravity in $d+2$-dimensions}

\label{sect:AA} We consider $d+2$-dimensional (with $d\geq 2$) Einstein
gravity minimally coupled to a scalar field $\phi $, 
\begin{equation}
I=\int d^{d+2}x\sqrt{-g}\left[ R-2(\partial \phi )^{2}-V\left( \phi \right) %
\right] .  \label{a1}
\end{equation}%
We will focus on models for which the scalar self-interaction potential $%
V(\phi )$ is given by 
\begin{equation}
V(\phi )=-\frac{d(d+1)}{\gamma L^{2}}\left( e^{2s\beta \phi }-\beta
^{2}e^{2\frac{s}{\beta }\phi }\right) \text{\ \ \ , \ \ \ }\gamma
=1-\beta ^{2},\quad {s}=\sqrt{\frac{2(d+1)}{d}},  \label{b1}
\end{equation}%
where $\beta $ is a (real) parameter characterizing the model and $L$ is the
AdS length. 
The action (\ref{a1}) is the $d+2$-dimensional generalization of
the four-dimensional Einstein-scalar gravity model investigated in Refs. 
\cite{Cadoni:2011yj,Cadoni:2012uf}. It shares with the 4D model of Refs 
\cite{Cadoni:2012uf,Cadoni:2011yj} 
several interesting
features. The potential (\ref{b1}) has a minimum at $\phi =0$ with $%
V(0)=-d(d+1)/L^{2}$, corresponding to an AdS$_{d+2}$ vacuum and a local
scalar excitation of positive squared-mass $m^{2}=2(d+1)^{2}/L^{2}$. 
The model is a fake SUGRA model. In fact the
potential (\ref{b1}) can be derived from the superpotential 
$P(\phi )=\sqrt{\frac{d}{2}}\gamma ^{-1}L^{-1}\left( e^{s\beta \phi
}-\beta ^{2}e^{\frac{s}{\beta }\phi }\right).$
The action (\ref{a1}) is invariant under the duality transformation 
\begin{equation}
\beta \rightarrow \frac{1}{\beta }.  \label{c1}
\end{equation}%

\subsection{Black brane solutions}

We now look for static, radially symmetric, planar solutions of the field
equations stemming from the action (\ref{a1}).

The presence of the $\phi =0$ minimum of the potential (\ref{b1}) for every
value of $\beta $, implies the existence of the Schwarzschild-AdS 
(SAdS) solution with $\phi=0$:
\begin{equation}
ds^{2}=-f(r)dt^{2}+f^{-1}(r)dr^{2}+r^{2}dx_{i}dx_{i}\ ,\qquad f(r)=\frac{%
r^{2}}{L^{2}}-\frac{2M}{r^{d-1}},  \label{SADS}
\end{equation}%
where $M$ is the black brane mass and $i=1,2\ldots d$. 

Solutions with a
non-trivial scalar field can be found using the procedure used in Ref. 
\cite{Cadoni:2012uf,Cadoni:2011yj} for the 4D case. Using for the metric the parametrization, 
\begin{equation}
ds^{2}=-e^{2\nu }dt^{2}+e^{2\nu +2d\rho }d\xi ^{2}+e^{2\rho }dx_{i}dx_{i},
\label{met}
\end{equation}%
the field equations can be recast in the form of the $SU(2)\times SU(2)$
Toda molecule \cite{Olshanetsky:1981dk}. Solutions with a
regular horizon and nontrivial scalar field exist only if they do not
approach asymptotically to AdS$_{{d+2}}$. The form of these hairy black
brane solutions can be found using the procedure explained in Ref. 
\cite{Cadoni:2012uf,Cadoni:2011yj}.
For $\beta ^{2}<1$ we get the two-parameter family of solutions: 
\begin{eqnarray}
ds^{2} &=&\left( \frac{r_{0}}{r}\right) ^{\frac{2}{\omega }}\left\{ \Delta
(r)^{\frac{2\beta ^{2}}{(d+1)\gamma }}\left[ -\Gamma \left( r\right)
dt^{2}+dx_{i}dx_{i}\right] +E\Delta (r)^{\frac{2\beta ^{2}}{\gamma }}\Gamma
\left( r\right) ^{-1}dr^{2}\right\} ,  \notag \\
e^{2\phi } &=&\left[ \frac{A}{\Delta (r)}\right] ^{\frac{2\beta }{s%
\gamma }}\left( \frac{r}{r_{0}}\right) ^{\frac{ds\beta }{\omega }%
},\quad \Gamma (r)=1-\mu _{1}\left( \frac{r_{0}}{r}\right) ^{\delta },\quad
\Delta (r)=1+\mu _{2}\left( \frac{r_{0}}{r}\right) ^{\delta },  \notag \\
\omega &=&1-(d+1)\beta ^{2},\quad \delta =-\frac{(d+1)\gamma }{\omega }%
,\quad A=\sqrt{\mu _{2}(\mu _{1}+\mu _{2})},\quad E=\left( \frac{\gamma L}{%
r_{0}\omega }\right) ^{2}A^{-\frac{2\beta ^{2}}{\gamma }},  \label{sol}
\end{eqnarray}%
where $\mu _{1,2}$ are dimensionless free parameters and $r_{0}$ is a length
scale that must be introduced in order to get the correct physical
dimensions.

The asymptotic region of the spacetime (\ref{sol}) is given by $r\rightarrow
0$ for $\beta ^{2}<1/(d+1)$, whereas it is given by $r=\infty $ when $\beta
^{2}\geq 1/(d+1)$. In both cases the asymptotic behavior of the solution (%
\ref{sol}) is given by 
\begin{equation}
ds^{2}=\left( \frac{r_{0}}{r}\right) ^{\frac{2}{\omega }%
}(-dt^{2}+dx_{i}dx_{i}+dr^{2}),\quad \phi =\frac{s d\beta }{2\omega }%
\log (r/r_{0}).  \label{e1}
\end{equation}%
This metric is not invariant under scale transformation, but still
transforms with definite weight, so it is scale-covariant.

The solution (\ref{sol}) becomes singular for $\beta ^{2}=1/(d+1)$. This is
related to the fact that this value of $\beta $ corresponds to a divergent
hyperscaling parameter $\theta $. Nevertheless a fully regular solution can
be written using a different parametrization for the radial coordinate $r$, 
\begin{eqnarray}
ds^{2} &=&\left( \frac{r_{0}}{r}\right) ^{2}\left\{ \Delta (r)^{\frac{2}{%
d(d+1)}}\left[ -\Gamma \left( r\right) dt^{2}+dx_{i}dx_{i}\right] +E\Delta
(r)^{\frac{2}{d}}\Gamma \left( r\right) ^{-1}\left( \frac{r_{0}}{r}\right)
^{2}dr^{2}\right\} ,  \notag  \label{k1a} \\
e^{2\phi } &=&\left[ \frac{A}{\Delta (r)}\right] ^{\sqrt{\frac{2}{d}}}\left( 
\frac{r}{r_{0}}\right) ^{\sqrt{2d}},\quad \delta =-d\quad E=\left( \frac{%
d^{2}\ L}{r_{0}(d+1)(d+2)}\right) ^{2}A^{-\frac{2}{d}},
\end{eqnarray}%
whereas $\Delta ,\Gamma ,A$ are given as in Eq. (\ref{sol})

The radial coordinate in the metric (\ref{e1}) gives the information about
the various energy scales in the dual QFT. A proper energy ${\mathcal{E}}%
_{0} $ is redshifted according to the law: 
\begin{equation}
{\mathcal{E}}(r)=r^{-\frac{1}{\omega} }{\mathcal{E}}_{0}.  \label{n1}
\end{equation}%
This equation tells us that for $\omega >0$ ($\beta ^{2}<1/(d+1)$), $%
r\rightarrow 0$ ($\rightarrow \infty $) corresponds to the UV (IR) region of
the dual QFT, whereas for $\omega <0$ ($\beta ^{2}>1/(d+1)$) the UV (IR)
corresponds to $r=\infty $ ($r\rightarrow 0$).

For $\mu _{1},\mu _{2}\geq 0$ , the metric (\ref{sol}) exhibits a
singularity at $r=\infty$ ($r=0 $) for $\beta ^{2}<1/(d+1)$ ($\beta ^{2}\geq
1/(d+1)$) shielded by a horizon at $r/r_{0}=\mu _{1}^{\,1/\delta }$, and
therefore represents a regular black brane.

Until now we have considered only the case $\beta ^{2}<1$. The form of the
solutions for $\beta ^{2}>1$ can be simply found using the duality (\ref{c1}%
) into Eq. (\ref{sol}). All the considerations of this section can be
trivially extended to the case $\beta ^{2}>1.$

\subsection{Extremal limit and scalar soliton}

The extremal limit of the solution (\ref{sol}) is obtained setting $\mu
_{1}=0$. When $\mu _{2}=0$ this extremal limit is singular, with a naked
singularity at $r=\infty $ for $\beta ^{2}<1/(d+1)$ (at $r=0$ for $\beta
^{2}>1/(d+1)$) with $\phi \sim \ln r$. On the other hand for $\mu _{2}>0$
the extremal BB is a regular scalar soliton that interpolates between a scale covariant  
solution in the UV and the AdS$_{d+1}$ in the IR: 
\begin{eqnarray}
ds^{2} &=&\left( \frac{r_{0}}{r}\right) ^{\frac{2}{\omega }}\left\{ \Delta
(r)^{\frac{2\beta ^{2}}{(d+1)\gamma }}\left[ -dt^{2}+dx_{i}dx_{i}\right]
+E\Delta (r)^{\frac{2\beta ^{2}}{\gamma }}dr^{2}\right\} ,  \notag
\label{f1} \\
e^{2\phi } &=&\left[ \frac{\mu _{2}}{\Delta (r)}\right] ^{\frac{2\beta }{%
s\gamma }}\left( \frac{r_{0}}{r}\right) ^{-\frac{ds\beta }{%
\omega }}.
\end{eqnarray}%
Let us now consider the UV (asymptotic) and IR (near-horizon) limit of the
scalar soliton (\ref{f1}). For $\beta ^{2}<1/(d+1)$ this corresponds to
take, respectively, $r\rightarrow 0$ and $r\rightarrow \infty $. For $\beta
^{2}\geq 1/(d+1)$ these limits are reversed (the UV corresponds to $%
r\rightarrow \infty $ and the IR to $r\rightarrow 0$).

In the IR limit, the scalar field $\phi $ vanishes, the length
scale $r_{0}$ decouples and the metric (\ref{f1}) becomes that of  
AdS$_{d+2}$. The length-scale $r_{0}$ is an UV scale, which decouples in the IR,
where conformal invariance is restored. On the other hand, in the UV limit
it is the AdS length $L$ that decouples: the metric (\ref{f1}) can be
written in terms of $r_{0}$ only and takes the scale-covariant form given by
Eq. (\ref{e1}).

\section{ Compactifications of $p$-brane solutions of SUGRA theories}

\label{sect:AB} In this section we will look for string theory realizations
that produce, after compactification, an Einstein-scalar model (\ref{a1})
with potential of the form (\ref{b1}). This means that we are considering
our models just as as effective description, which breaks down in the far UV. The
short-distance physics will be therefore described by the UV completion of
our effective model.

We will show in the following that BB solutions (\ref{sol}) arise as
compactifications of black $p$-brane solutions of SUGRA theories. We will
see that they emerge from the $p$-brane both as simple spherical
compactification or also as a more general Kaluza-Klein compactification
parametrized by a  parameter. 

Black $p$-branes are classical
Ramond-Ramond charged solutions of $D$-dimensional SUGRA theories supported
by a $(p+2)$-form field strength $G_{p+2}$ \cite{Horowitz:1991cd,Duff:1994an}.
In the Einstein frame
the bosonic part of the action is 
\begin{equation}
I=\int d^{D}x\,\sqrt{-g}\left( R-\frac{1}{2}(\partial \Phi )^{2}-e^{a\Phi }%
\frac{1}{2(p+2)!}G_{(p+2)}\right) ,
\end{equation}%
where $\Phi $ is the dilaton field and $a$ is constant, which is zero for
non-dilatonic $p$-branes, whereas 
\begin{equation}
a^{2}=4-[(p+1)(D-p-3)]/(D-2),  \label{t1}
\end{equation}%
for dilatonic branes. The metric part of the $p$-brane solution is given in
terms of two integration constants $h_{0},g_{0}$ by 
\begin{eqnarray}
ds_{D}^{2} &=&H(r)^{-\frac{2\tilde{d}}{\rho }}\left(
-g(r)dt^{2}+\sum_{i=1}^{p}dx_{i}dx_{i}\right) +H(r)^{\frac{2{d}}{\rho }%
}\left( g(r)^{-1}dr^{2}+r^{2}d\Omega ^{2}_{q}\right) ,  \notag  \label{j8} \\
H(r) &=&1+\left( \frac{h_{0}}{r}\right) ^{\tilde{d}},\,\,g(r)=1-\left( \frac{%
g_{0}}{r}\right) ^{\tilde{d}},\,\,\rho =(p+1)\tilde{d}+a^{2}\frac{D-2}{2}%
,\,\,\tilde{d}=D-p-3,
\end{eqnarray}%
where $d\Omega ^{2}_{q}$ is the line element of a compact space ${\mathcal{%
K}}^{q}$ with $q=D-p-2$ dimensions.

Let us first consider nondilatonic $p$-branes. 
The simplest diagonal ansatz for the D-dimensional metric,
which gives the $p+2$-dimensional theory in the Einstein frame, is  
obtained by setting ${\mathcal{%
K}}^{q}=S^{q}$ and 
\begin{equation}
ds_{D}^{2}=e^{-\frac{2q}{p}\psi }ds_{p+2}^{2}+e^{2\psi }d\Omega ^{2}_{q}.
\label{k8}
\end{equation}%
Taking into account that for nondilatonic branes $e^{\psi }=rH^{1/ \tilde d}$
one finds after compactification the BB metric: 
\begin{equation}
ds_{p+2}^{2}=r^{-\frac{2}{p}(p+2-D)}\left[ H(r)^{\frac{2(D-2)}{p(p+1)(D-p-3)}%
}\left( -g(r)dt^{2}+\sum_{i=1}^{p}dx_{i}dx_{i}\right) +H(r)^{\frac{2(D-2)}{%
p(D-p-3)}}g(r)^{-1}dr^{2}\right] .  \label{j9}
\end{equation}%
One can easily see that the metric (\ref{j9}) matches exactly, after some
trivial identification of the parameters, the metric (\ref{sol}) if we take $%
d=p$ and 
\begin{equation}
\beta ^{2}=\frac{D-2}{(p+1)(D-p-2)}.  \label{l1}
\end{equation}%
It is important to notice that this value of $\beta $ always satisfy the
inequality $1/(p+1)<\beta ^{2}<1$. Particularly interesting cases are
represented by the 2 and 5-brane in $D=11$ corresponding, 
respectively, to $%
\beta ^{2}=3/7$ and $\beta ^{2}=3/8$.

Compactification of $p$-branes with the diagonal ansatz (\ref{k8}) produces
BB solutions of the form (\ref{sol}) only for the particular values of the
parameter $\beta $ given in Eq. (\ref{l1}). This limitation can be removed
by considering the more general diagonal ansatz of Ref. \cite{Gouteraux:2011ce} for the
D-dimensional metric.

Let us now briefly consider compactification of dilatonic $p$-branes. In
this case we must use in (\ref{j8}) the value (\ref{t1}) for $a$ giving $%
\rho =2(D-2)$. The diagonal ansatz (\ref{k8}) produces now the BB solution 
\begin{equation}
ds_{p+2}^{2}=r^{-\frac{2}{p}(p+2-D)}\left[ H(r)^{\frac{1}{p}}\left(
-g(r)dt^{2}+\sum_{i=1}^{p}dx_{i}dx_{i}\right) +H(r)^{\frac{p+1}{p}%
}g(r)^{-1}\ dr^{2}\right] .  \label{j10}
\end{equation}%
Matching this BB solution with  Eq. (\ref{sol}) requires 
\begin{equation}
D=\frac{3p+1}{p-1}+p+2,\quad \beta ^{2}=\frac{p+1}{3p+1}.  \label{j11}
\end{equation}%
These are very stringent constraints which however are satisfied by a very
interesting case, the 3-brane in $D=10$ which gives $\beta ^{2}=2/5$. It is
likely that also for dilatonic branes the use of the more general diagonal
ansatz of Ref. \cite{Gouteraux:2011ce} would allow to circumvent the constraints (\ref{j11}%
). In this paper we will not further investigate on this point.

\section{Thermodynamics and phase transitions}

\label{sect:AC} In this section we will consider the BB solutions (\ref{sol}%
) as a thermodynamical system. We will use the euclidean action formulation
of Martinez et al.\ \cite{Martinez:2004nb}. As it has been already noted in
Ref. \cite{Cadoni:2012uf} for the 4D case, the two-parameter family of solutions (\ref%
{sol}) is not suitable for setting up a consistent BB thermodynamics. The
problem is the explicit dependence of the scalar field from the parameter $%
\mu _{1}$, which causes divergences in the boundary action, that determines
the mass of the solution. This explicit dependence can be eliminated 
by constraining the possible values of $\mu _{1,2}$ in
Eq. (\ref{sol}) with $\mu _{2}(\mu _{2}+\mu _{1})=1$. We end
up with the one-parameter family of solutions 
\begin{eqnarray}
ds^{2} &=&r^{-\frac{2}{\omega }}\left\{ \left( \frac{\gamma L}{\omega }%
\right) ^{2}\left[ -\Delta (r)^{\frac{2\beta ^{2}}{(d+1)\gamma }}\Gamma
(r)dt^{2}+\Delta (r)^{\frac{2\beta ^{2}}{\gamma }}\Gamma (r)^{-1}dr^{2}%
\right] +\Delta (r)^{\frac{2\beta ^{2}}{(d+1)\gamma }}dx_{i}dx_{i}\right\} ,
\notag \\
e^{2\phi } &=&\Delta (r)^{-\frac{2\beta }{s\gamma }}r^{\frac{s%
d\beta }{\omega }},\quad \Gamma (r)=1-\frac{\nu _{1}}{r^{\delta }},\quad
\Delta (r)=1+\frac{\nu _{2}}{r^{\delta }},  \label{f3}
\end{eqnarray}%
where the parameters $\nu _{12}$ are constrained by 
\begin{equation}
\nu _{1}=\frac{1}{\nu _{2}}-\nu _{2},\quad 0<\nu _{2}\leq 1,\quad 0\leq \nu
_{1}<\infty  \label{constraint}
\end{equation}%
Notice that in writing Eq. (\ref{f3}) we have introduced dimensionless
coordinates $t,r$ and parameters $\nu _{12}$. This is necessary because (\ref%
{f1}) is a global solution interpolating between the IR and the UV 
regimes
that are characterized by two different length scales $r_{0}$ and $L$.

Starting from Eq. (\ref{f3}) one can now calculate, using standard formulas,
the temperature $T$ and entropy $S$ of the BB. One has 
\begin{equation}
T=\frac{1}{4\pi }\frac{d(d+1)\gamma }{d+2(d+1)\beta ^{2}}\nu _{2}^{-\frac{%
\omega }{(d+1)\gamma }}(1-\nu _{2}^{2})^{1/(d+1)},\quad S=4\pi V\nu _{2}^{-%
\frac{d}{(d+1)\gamma }}(1-\nu _{2}^{2})^{\frac{d}{d+1}},  \label{k1}
\end{equation}%
where $V$ is the volume of the transverse $d$-dimensional space. 

We construct the  thermodynamics of our BB solutions 
using the Euclidean action formalism.
Thermodynamical potentials are given 
by boundary terms of the action. 
We use the  parametrization of the
metric of Ref. \cite{Martinez:2004nb}: 
\begin{equation*}
ds^{2}=N^{2}f^{2}dt^{2}+f^{-2}dr^{2}+R^{2}dx_{i}dx_{i},
\end{equation*}%
{\textbf{\ }}The variations of the boundary terms  of the action 
evaluated  for the solution
(\ref{f3}) are

\begin{eqnarray}
\delta I_{G}^{\infty } &=&-\frac{Vd^{2}}{T\left[ (d+2(d+1)\beta ^{2})\right] 
}\left[ \delta \nu _{1}+\frac{2\beta ^{2}}{\gamma }(2-\beta ^{2})\delta \nu
_{2}\right] ,  \notag  \label{g1} \\
\delta I_{\phi }^{\infty } &=&\frac{2\beta ^{2}Vd^{2}}{\gamma T\left[
(d+2(d+1)\beta ^{2})\right] }\delta \nu _{2}, \\
\left. \delta I_{G}\right\vert _{r_{h}} &=&-\frac{Vd^{2}}{T\left[
(d+2(d+1)\beta ^{2})\right] }\left( \nu _{1}+\nu _{2}\right) ^{-1}\left[
(\nu _{1}+\gamma \nu _{2})\delta \nu _{1}+\beta ^{2}\nu _{1}\delta \nu _{2}%
\right] ,  \notag \\
\left. \delta I_{\phi }\right\vert _{r_{h}} &=&0.  \notag
\end{eqnarray}
where $I_{G}$ and $I_{\phi }$ are, respectively, the gravitational and scalar
field part of the boundary action.

One can easily show that the BB entropy $S$ is correctly given by $S=\left.
I_{G}\right\vert _{r_{h}}$. The mass of the BB is given in terms of the free
energy $F$ and the entropy $S$ by $M=F+TS=-T(I_{G}^{\infty }+I_{\phi
}^{\infty })$. Using Eqs.\ (\ref{g1}) one finds 
\begin{equation}
M=\frac{Vd^{2}}{d+2(d+1)\beta ^{2}}\left( \nu _{1}+2\beta ^{2}\nu
_{2}\right) =\frac{Vd^{2}}{d+2(d+1)\beta ^{2}}\left\{ \frac{1}{\nu _{2}}%
+\left( 2\beta ^{2}-1\right) \nu _{2}\right\} .  \label{mass}
\end{equation}%
Using Eqs. (\ref{k1}), (\ref{mass}) and the constraint (\ref{constraint})
one can now check that the first principle $dM=TdS$ is satisfied. 
As usual the results can
be trivially extended to the parameter region $\beta ^{2}>1$ just by using
the duality $\beta \rightarrow 1/\beta $ in Eqs. (\ref{f3}), (\ref{k1}) and (%
\ref{mass}).

\subsection{Phase transition}

The global stability of our BB solution, considered as a thermodynamical
system, can be investigated by computing the free energy and the specific
heat. In particular, comparison of the free energies of different
configurations at fixed temperature allows us to single out the
energetically preferred phase, whereas a positive (negative) specific heat
indicates local stability (instability) of a given phase. We start with the
case $\beta ^{2}<1/(d+1)$, where, as we will see, we observe a phase
transition.

\subsubsection{Free energy}

In the case under consideration the two competitive phases are represented
by the black brane with scalar hair (scalar BB) (\ref{f3}) and the SAdS BB (%
\ref{SADS}). The free energy of the scalar black brane is 
\begin{equation}
F_{SB}(T)=M-TS=\frac{Vd}{d+2(d+1)\beta ^{2}}\left\{ -\frac{\omega }{\nu
_{2}(T)}+\left[ 1+(d-1)\beta ^{2}\right] \nu _{2}(T)\right\} ,  \label{free}
\end{equation}%
where $\nu _{2}(T)$ is defined implicitly by the first equation in (\ref{k1}%
). For the free energy of the SAdS black brane we have instead, 
\begin{equation*}
F_{SAdS}(T)=-V\left( \frac{4\pi }{d+1}\right) ^{d+1}T^{d+1}.
\end{equation*}%
The relevant quantity $\Delta F(T)=F_{SB}(T)-F_{SAdS}(T)$ cannot be computed
explicitly in closed form because $\nu _{2}(T)$ is only implicitly defined.
Nevertheless, one can show that for $\beta ^{2}<1/(d+1)$, $\Delta F(T)$ is
positive for small $T$, vanishes at finite value of the temperature and
becomes negative at large $T$.

A qualitative way to see this change of sign of $\Delta F(T)$ is to consider
the small-$T$ ($\nu _{2}\sim 1$) and the large-T ($\nu _{2}\sim 0$) behavior
of $F_{SB}$. At small temperatures we have 
\begin{equation}
F_{SB}(T)=V\left\{ \frac{2\beta ^{2}d^{2}}{d+2(d+1)\beta ^{2}}-\frac{\left(
4\pi \right) ^{d+1}}{(d+1)^{d+1}}\left[ \frac{d+2(d+1)\beta ^{2}}{d\gamma }%
\right] ^{d}T^{d+1}\right\} .  \label{g5}
\end{equation}%
The small-$T$ behavior is determined  by the $T=0$, AdS$_{d+2}$
extremal limit and is pertinent  to a holographically dual $(d+1)$-dimensional
CFT. Conversely for the
large-$T$ ($\nu _{2}\sim 0$) behavior we have
\begin{equation}
F_{SB}=-\frac{\omega V d}{d+2(d+1)\beta ^{2}}\left\{ \frac{4\pi \left[
d+2(d+1)\beta ^{2}\right] }{d(d+1)\gamma }\right\} ^{\frac{(d+1)\gamma }{%
\omega }}T^{\frac{(d+1)\gamma }{\omega }}.  \label{lib}
\end{equation}

The free energy  for the hairy BB is positive at small $T$ implying $\Delta F>0$.
For $\beta ^{2}<1/(d+1)$, $\Delta F$ becomes negative at large $T$. 
This shows  the existence of a
critical temperature $T_{c}$ such that $\Delta F(T_{c})=0$.

In general the critical temperature cannot  be determined 
analytically.  However, we can  show the existence of $T_{c}$ graphically. By
setting $y=\nu _{2}^{2}$ the equation $F_{SB}=F_{SAdS}$ gives: 
\begin{equation*}
g(y)=\frac{1-(d+1)\beta ^{2}-\left[ 1+(d-1)\beta ^{2}\right] y}{1-y}=f(y)=%
\left[ \frac{d}{d+2(d+1)\beta ^{2}}\right] ^{d}\gamma ^{d+1}y^{\frac{d\beta
^{2}}{2\gamma }},\text{ \ \ \ \ \ }0\leq y\leq 1.
\end{equation*}%
While the curve $f(y)$ is always positive, the behavior of $g(y)$ depends on
the value of $\beta $. For $\beta ^{2}>\frac{1}{d+1}$, the curve starts from
a negative point and is always negative; for $\beta ^{2}=\frac{1}{d+1}$, the
curve starts from $y=0$ and is always negative; for $\beta ^{2}<\frac{1}{d+1}
$, the curve starts from a positive point and decreases monotonically to $%
-\infty $. Then the two curves $f(y)$ and $g(y)$ do not intersect for $\beta
^{2}\geqslant \frac{1}{d+1}$, while they intersect at a finite critical
value of the temperature for $\beta ^{2}<\frac{1}{d+1}$.

\begin{figure}[th]
\begin{center}
\begin{tabular}{cc}
\epsfig{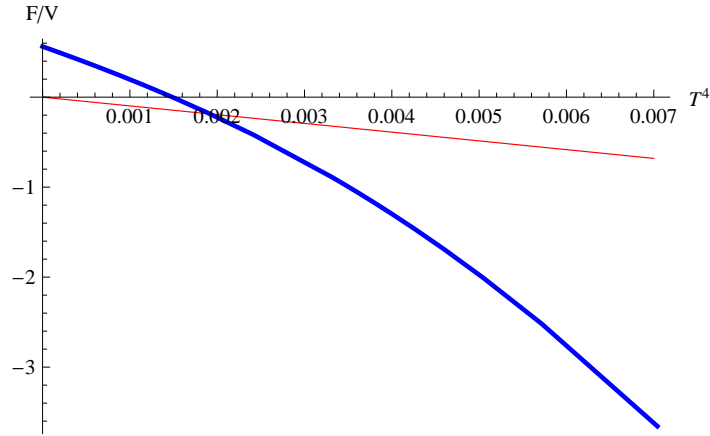} & 
\end{tabular}%
\end{center}
\caption{The free energy density $F/V,$ as function of $T^{4},$ of the
scalar black brane for $d=3$ and $\protect\beta ^{2}=1/8$ (blue, thick line)
and of the SAdS black brane for $d=3$ (red, thin line). }
\label{fig1}
\end{figure}

In figure \ref{fig1} we show the behavior of the free energy density 
for $d=3$
and for $\beta ^{2}=1/8$, a value  in the range $0\leq
\beta ^{2}<1/(d+1)$. The critical temperature can also be determined 
numerically. For the case described in Fig. \ref{fig1} we have $T_{c}=0.20917$.

\bigskip

We have therefore discovered, in the  $\beta ^{2}<1/(d+1)$ case,
a cross-over behaviour  for the free energies of 
the SAdS and scalar black brane solution.  The relevant question is 
now the following: can we interpret this behaviour as a phase 
transition between two different configuration of the same bulk 
gravity theory?  This question can be answered only if one clarifies 
the role played by boundary conditions in the definition of  
canonical  thermodynamical ensembles. In fact, the two classical 
configurations - the SAdS and the scalar brane - actually are two 
different solutions of the same bulk theory defined by the action    
(\ref{a1}).  On the other hand,  these solutions correspond to  different asymptotic 
values 
of the scalar field in the UV ($\phi 
= 0$ and $\phi=-\infty$ respectively  for the SAdS 
solution (\ref{SADS}) and the hairy black brane solution (\ref{sol}) ) 
and to different asymptotic geometries. 
However, there is no obstruction in considering solutions of the same 
bulk theory  with 
different boundary conditions as belonging to the same canonical 
ensemble. Although this is not an usual situation in the AdS/CFT 
correspondence, where one refers to fixed boundary conditions, 
one can define a canonical partition function just by 
evaluating the euclidean action on the particular bulk solution, 
without any reference to the asymptotics of the solutions.  This is 
exactly the way we have   calculated  the free energy using Eq. 
(\ref{g1}). 
This is a strong argument supporting the interpretation of the free 
energy cross-over described in this section as a truly first-order 
phase transition (the phase transition is
first-order because at $T=T_{c}$, $dF_{SB}/dT\neq dF_{SAdS}/dT$).

A definitive answer about the existence of the phase transition could be 
obtained by showing that for $T>T_{C}$ the SAdS solution decays with 
finite half-life in the scalar brane solution, but such a 
calculation is beyond the scope of this paper.

Because of the change in asymptotics in the  two competitive bulk 
solutions, the holographic interpretation of the phase transition is rather 
involved.  
In the usual gravity/gauge theory  correspondence
dictionary  the sources $J$ in the dual QFT are related to small 
perturbations of the UV boundary conditions. The gravity/gauge theory  correspondence
rules allow then to compute the $n$-point functions for dual operators 
differentiating the bulk partition function with respect to $J$.
The dramatic change in the boundary conditions for the scalar field 
$\phi$  we have in our case seems to suggest that the two different 
phases we have on the gravity side correspond to different sources  
in the dual QFT.  Because different sources
generally lead to different Lagrangians, we are led to the conclusion
that the two phases of the gravity theory - the SAdS and the scalar 
brane phase- correspond to two distinct dual  QFTs, not  to two
distinct phases of a single QFT\footnote{We thank the anonymous 
referee of JHEP for suggesting to us this interpretation.}.

An other argument supporting this interpretation is the analogy with 
what happens in bulk theories allowing for a flow between an AdS in 
the IR and an other AdS in the UV. Such solutions are known in the 
literature.  Analogously to the case discussed in this paper, we have 
also here three 
different boundary QFTs. We have  two CFTs with no flow, corresponding to fixed 
IR or UV fixed values for the scalar field and a  QFT describing the flow 
between the IR and the UV fixed point, corresponding to a 
$r$-dependent scalar field.

\subsubsection{Specific heat}

It is easy to check, using Eqs. (\ref{k1}) and (\ref{mass}), that for $\beta
^{2}<1/(d+1)$ the function $M(T)$ is a monotonic increasing function of the
temperature $T.$ Then the specific heat $c=\partial M/\partial T$ is
positive for all values of $T.$ Similarly, the specific heat of the
SAdS black brane is:
$c_{SAdS}(T)=\frac{(4\pi )^{d+1}Vd}{(d+1)^{d}}T^{d}>0$.

\subsection{$1/(d+1)\leq \protect\beta ^{2}<1$}

For $\beta ^{2}=1/(d+1)$, the scalar black brane solution exists only for
temperatures below the critical value $T=T_{c}=\frac{d^{2}}{4\pi (d+2)}$,
while for $T>T_{c}$ only the SAdS solution (\ref{SADS}) exists. The free
energy is: 
\begin{equation*}
F_{SB}=\frac{2Vd^{2}}{(d+1)(d+2)}\sqrt{1-\left( \frac{T}{T_{c}}\right) ^{d+1}%
}.\text{ \ \ \ \ \ }
\end{equation*}%
The free energy is positive definite and vanishes for $T=T_{c}$, while 
$F_{SAdS}$ is always negative. Then we have $F_{SAdS}<F_{SB}$ in the whole
range $T\leq T_{c}$, that is the SAdS solution is always energetically
favored. The specific heat of the black brane solution is always positive and
diverges at the critical temperature.

For $\beta ^{2}>1/(d+1)$, the function $T(\nu _{2})$ is not  
monotonic. It has a maximum at $\nu _{2}=\nu _{0}=\sqrt{\left[
(d+1)\beta ^{2}-1\right] /\left[ (d-1)\beta ^{2}+1\right] }$. Also in this
case the black brane solution exists only below a maximum, critical
temperature $T=T_{c}$.

For what concerns the free energy, from Eq. (\ref{free}) it is easy to
realize that, for $\beta ^{2}>1/(d+1)$, $F_{SB}$ is always positive. Hence
also in this case $F_{SAdS}<F_{SB}$ and the the SAdS solution is
energetically preferred with respect to the scalar-dressed black brane.
However, the non-monotonicity of the function $T(\nu _{2})$ implies the
existence of two different branches of the SB phase for $T\leq T_{c}$, as it
has been already observed in Ref. \cite{Cadoni:2012uf} for the 4D case. The first branch
(obtained for $\nu _{0}\leq \nu _{2}\leq 1$) is the analogue of the $%
AdS_{d+2}$ phase obtained for $\beta ^{2}<1/(d+1)$ at small $T$, while the
second branch (obtained for $0<\nu _{2}\leq \nu _{0}$) has no analogue for $%
\beta ^{2}<1/(d+1)$. In this case the free energy scales at small
temperature as $F\sim T^{\alpha }$, with $\alpha =(d+1)\gamma /\omega $. But 
$\alpha $ is negative, hence $F_{SB}$ has a singularity at $T=0$. 

For what concerns the specific heat we have an interesting peculiarity: 
in the first branch  the specific heat  is
positive and hence it corresponds to a locally stable phase (although this
phase is not energetically preferred with respect to the SAdS solution),
while in the second branch 
$c(T)$ is always negative, corresponding to an unstable phase.

\subsection{Dual solution}

As already observed, using the duality (\ref{c1}) into the (\ref{f3}) we
obtain the solution for $\beta ^{2}>1$. 
The thermodynamical properties of these solutions follow easily from the
case $\beta ^{2}<1$ by duality. We note that in this case the phase
transition between the scalar-dressed black brane solution and the SAdS
solution is present for $\beta ^{2}>(d+1)$, while for $\beta ^{2}\leq (d+1)$
the SAdS solution is always energetically favored respect to the SB
solution. The behaviors of the free energy and the specific heat in the
three cases are qualitatively analogous to those discussed for $\beta ^{2}<1$.

\bigskip

\bigskip

\bigskip

{\textbf{\ }}

\section{Hyperscaling violation}

\label{sect:AD} 

The   thermodynamical behavior    of our  scalar BB 
described in the previous sections is strongly related to
the symmetries of the solutions 
in the UV and IR regimes.

The UV regime, where the solution
takes the form (\ref{e1}), is characterized  by 
violation of the scale symmetry, whereas in  the IR regime 
we have the conformal invariant AdS$%
_{d+2}$ extremal solution. For the  dual QFT this
 corresponds to a hyperscaling-violating phase in the UV and to a
scaling-preserving phase in the IR. 

To describe holographic hyperscaling violation in $d+2$ dimensions we use
the parametrization of the the scale covariant metric of Ref. \cite%
{Dong:2012se}: 
\begin{equation}
ds^{2}={r^{-2(d-\theta )/d}}\left( -{r^{-2(z-1)}}dt^{2}+dx_{i}dx_{i}+dr^{2}%
\right) ,  \label{hv}
\end{equation}%
where $\theta $ is the hyperscaling violation parameter and $z$ is the
dynamic critical exponent (it describes anisotropic scaling, hence violation
of Poincar\'{e} symmetry, in the $(d+1)$-dimensional spacetime). The
transformation law under rescaling of the coordinates is 
\begin{equation}
t\rightarrow \lambda ^{z}t,\quad x_{i}\rightarrow \lambda x_{i},\quad
r\rightarrow \lambda r,\quad ds\rightarrow \lambda ^{\theta /d}ds.
\label{tl}
\end{equation}

The
scaling transformation (\ref{tl}) determines the following scaling behavior for
the free energy: 
\begin{equation}
F\sim T^{\frac{(d-\theta )+z}{z}}.  \label{scal}
\end{equation}%
This relation allows a simple physical interpretation of the hyperscaling
violation parameter $\theta $ in terms of the hyperscaling relation between
specific heat exponent $\hat{\alpha}$ and critical exponent $\hat{\nu}$. The
relation $2-\hat{\alpha}=d\,\hat{\nu}$ is modified by \textquotedblleft
lowering\textquotedblright\ the dimensionality of the system from $d$ to $%
d-\theta $, namely $2-\hat{\alpha}=(d-\theta )\hat{\nu}$.

Comparing Eq. (\ref{hv}) with Eq. (\ref{e1}) one can easily read off the
parameters $\theta ,z$ for our BB solution: 
\begin{equation}
z=1,\text{ \ \ }\theta =\frac{d(d+1)\beta ^{2}}{(d+1)\beta ^{2}-1}.
\label{espon}
\end{equation}%
As usual the case $\beta ^{2}>1$ is covered just by using the duality (\ref%
{c1}). We have 
\begin{equation}
z=1,\text{ \ \ }\theta =\frac{d(d+1)}{(d+1)-\beta ^{2}}.  \label{espon1}
\end{equation}

As expected, we have $z=1,\,\theta\neq 0$ in the scalar black brane phase, 
whereas we get $z=1,\,\theta=0$ in the SAdS phase. This
gives the deviation from the conformal scaling of the free energy of a $d+1$
conformal field theory. 

One can easily check from Eq. (\ref{espon}) that $\theta <0$ for $\beta ^{2}<%
\frac{1}{d+1}$ and $\theta >d$ for $\frac{1}{d+1}<\beta ^{2}<1$, while $%
\theta $ diverges{\textbf{\ }}for $\beta ^{2}=\frac{1}{d+1}$ (for the dual
case (\ref{espon1}) we have $\theta <0$ for $\beta ^{2}>d+1$ and $\theta >d$
for $\beta ^{2}<d+1$). The null energy conditions for the bulk stress-energy
tensor are satisfied: in fact for $z=1$ these conditions require either $%
\theta \leq 0$ or $\theta \geq d$ \cite{Dong:2012se}.

A negative value of $\theta $ is not common in condensed matter
critical system, for which $\theta$ is positive.
However in our solutions the case $\theta <0$ is physically more interesting
(in particular for the possible holographic applications) because in
this case we observe a phase transition between the scalar black brane
solution and the SAdS solution, and the specific heat of the BB solution is
always positive. 
\section{General models}

\label{sect:AEE}

In the previous sections we have investigated the
Einstein-scalar gravity model defined  by the potential (\ref{b1}).
However, the main features of our models  are
dictated not by the full form of the potential but only 
by the behavior of the potential at $\phi =0$ and $%
\phi =-\infty $. We will  show that the two main features of
the model (hyperscaling violation and the SAdS$\rightarrow $
scalar BB phase transition) are pertinent to all models satisfying the
conditions: 1) $V(\phi )$ has a local minimum for $\phi =0$ with $%
V(0)<0$; 2) The potential approaches zero exponentially as $\phi \rightarrow
-\infty $. 
The previous conditions ensure the existence of an AdS$_{d+1}$ vacuum and of
a Schwarzschild-AdS (SAdS) black brane solution with $\phi =0$. 

In Ref.\ \cite{Cadoni:2011nq} has been derived the general BB solution of a
model with an exponential potential in $d+2$ dimensions. In particular, for
the metric parametrization we are using in this paper, the asymptotic
behavior of the solutions for the exponential potential 
$V=-\l^{2}e^{2h \phi}$ is
given by 
\begin{equation}
\phi =-\frac{dh}{dh^{2}-2}\,\log r+\frac{1}{2h}\ln C_{1},\quad ds^{2}=r^{%
\frac{4}{dh^{2}-2}}\left( -dt^{2}+dx_{i}dx_{i}+dr^{2}\right) ,  \label{s2}
\end{equation}%
where $h>0$ and $C_{1}=\{2d[2(d+1)-dh^{2}]\}/[\lambda ^{2}(dh^{2}-2)^{2}]$.

The case $\beta ^{2}<1$ described in the previous section for the model (\ref%
{a1}) is covered by setting $h^{2}<2(d+1)/d$ , whereas the two cases $\beta
^{2}<1/(d+1)$ and $\beta ^{2}>1/(d+1)$ correspond, respectively, to $h^{2}<2/d$
and $h^{2}>2/d$.

For a generic model, the existence of
a global scalar black brane solution interpolating between the 
AdS$_{d+2}$ vacuum
and the asymptotic scale covariant  solution has to be shown  numerically. 
If we can prove that such a solution  exists, the thermodynamical
system for $h^{2}<2/d$ must have a 
scalar black brane $\rightarrow $ SAdS phase transition. 

The derivation follows closely that used in Sect. \ref{sect:AC}. At small $T$
the free energy of the scalar black brane must have a behavior similar to
that of Eq.\ (\ref{g5}), i.e.\ $F_{SB}=C_{2}-C_{3}T^{d+1},$ with $C_{2,3}$
positive constants. This implies that at small $T$, $F_{SB}-F_{SAdS}>0$. On
the other hand, at large $T$, the free energy scales as $F_{SB}\sim
-T^{(2+2d-dh^{2})/(2-dh^{2})}$. For $h^{2}<2/d$ we have $%
T^{(2+2d-dh^{2})/(2-dh^{2})}>T^{d+1}$, from which follows that at large $T$, 
$F_{SB}-F_{SAdS}<0$.  

Comparing Eq. (\ref{s2}) with Eq.\ (\ref{hv}), one can read off the
hyperscaling violation parameter and the dynamic critical exponent: 
\begin{equation}
\theta =\frac{d^{2}h^{2}}{dh^{2}-2},\quad z=1.  \label{g8}
\end{equation}%
Notice that $\theta $ is negative for $h^{2}<2/d$, whereas $\theta >d$ for $%
h^{2}>2/d$.

\section{Holographic properties and two-point functions for scalar operators}

\label{sect:AE}

Holographic features of theories with hyperscaling violation have been
discussed in Ref. \cite{Dong:2012se}. 
Most of the results derived in Ref. \cite{Dong:2012se} for general scale-covariant
metrics apply directly to the model discussed in this paper. Imposing on the
gravity side the null energy conditions on the stress-energy tensor
constrains the range of the possible values of the
parameters $z,\theta $. In our case, being $z=1$, the conditions of Ref. 
\cite{Dong:2012se} become simply $\theta \leq 0$ or $\theta >d$. Taking into account
Eq. (\ref{espon}) one can easily see that these conditions are always
satisfied for every value of $\beta $, being $\theta <0$ for $\{0<\beta
^{2}<1/(d+1)\}\cup \{d+1<\beta ^{2}<\infty \}$ and $\theta >d$ for $%
\{1/(d+1)<\beta ^{2}<d+1,\,\,\beta ^{2}\neq 1\}$.

In Ref. \cite{Dong:2012se} it has been also calculated the short distance form of the
two point function of a scalar operator $\mathcal{O}$ dual to a scalar field
with a potential $2m^{2}\phi ^{2}$. It has been shown that it has a
power-law form and for $z=1$, $0<\theta <d$ is given by \cite{Dong:2012se} 
\begin{equation}
\langle \mathcal{O}(x)\mathcal{O}(x^{\prime })\rangle =\frac{1}{|x-x^{\prime
}|^{2(d+1)-\theta }}.  \label{11}
\end{equation}%
The problem is that the derivation of Ref. \cite{Dong:2012se} does not hold for $%
\theta <0$, which is the most interesting case for the models under
consideration in this paper. Moreover, for $\theta >d$, $r\rightarrow 0$
corresponds to the IR regime of the dual QFT. This means that for $\theta >d$%
, Eq. (\ref{11}) gives the large distance behavior of the two-point function
instead of the short distance behavior.

Let us now first observe that for $\theta <0$  Eq. (\ref{11}) gives the
IR behavior of the two-point function.
This means that for $\theta <0$ the mass term is irrelevant in the IR and
dominates in the UV. Conversely, for $\theta >0$ we have the opposite
behavior: the mass term is irrelevant in the UV and becomes relevant in the
IR. It is exactly this feature that allows one to use scaling arguments to
determine the form (\ref{11}) for the two-point function.

Obviously, if the
theory whose solution is given by the metric (\ref{hv}) has an UV (or IR)
completion with an UV (or IR) fixed point, the far short (far large)
behavior of the two-point function (\ref{11}) will be modified accordingly.
This is for instance the case of the models discussed in this paper, which
have an IR fixed point.

We are therefore left with the problem of finding a short distance form for
two point functions of scalar operators in the case $\theta <0$. A strong
hint for tackling the problem can be obtained by looking at the
gravitational dynamics that produces solution (\ref{hv}). One can easily
realize that, at least in the context of Einstein-scalar gravity, what is
needed is an exponential potential and a $\ln r$ short distance behavior for
the scalar (see Eq. (\ref{s2})). We will therefore look for the UV behavior of 
two point function
of a scalar operator $\mathcal{O}$ dual to a scalar field that 
supports our black brane solution and therefore has  near the UV a potential $%
-\lambda ^{2}e^{2h\phi }$. The equation of motion for  $\phi $ in the
background (\ref{s2}) are  
\begin{equation}
\left( \partial _{r}^{2}-\frac{d-\theta }{r}\partial _{r}+\partial
_{i}^{2}-\partial _{t}^{2}\right) \phi +\frac{h\lambda ^{2}}{2}e^{2h\phi
}r^{-2+\frac{2\theta }{d}}=0,  \label{h6}
\end{equation}%
where $h$ has to be expressed as a function of $\theta $ using Eq. (\ref{g8}%
). Eq. (\ref{h6}) can be solved perturbatively for $\theta <0$ ($h^{2}<2/d$)
by expanding $\phi$ around the background solution $\phi_{0}$ given 
by  Eq. (\ref{s2}): $\phi=\phi_{0}+\delta \phi$.
Using Eqs. (\ref{s2}) and (\ref{g8}) one gets for the perturbation $
\delta \phi $, the equation of motion satisfied  by a massive scalar 
field in AdS in $d+2-\theta$ "bulk dimensions'' 
\footnote{We thank the anonymous referee of JHEP for pointing out to 
us this fact and for finding an error in the calculations 
 leading to Eq. (\ref{g9}) of  the 
previous version of this paper.}:
\begin{equation}
\left( \partial _{r}^{2}-\frac{d-\theta }{r}\partial _{r}+\partial
_{i}^{2}-\partial _{t}^{2}\right) \delta\phi 
-\frac{m^{2}}{r^{2}}\delta\phi=0 \label{h6a}
\end{equation}%
with $m^{2}=-C_{1}h^{2}\lambda^{2}=[2\theta(d+1-\theta)]/d$.
Eq. (\ref{h6a}) can be solved with the usual power-law ansatz 
$\delta\phi\propto r^{\alpha}(1+{\cal O}(r^{2}¥))$, with $\alpha$ given by the standard 
AdS formula in $d+1-\theta$ dimensions:%
\begin{equation}
\alpha _{12}=\frac{1}{2}\left(d+1-\theta \pm \sqrt{
\left( d+1-\theta \right)^{2}+4m^{2}}\right)=
\frac{1}{2} \left(d+1-\theta \right) 
\left( 1\pm \sqrt{1+ \frac{8\theta}{ d(d+1-\theta)}}\right). 
\label{g9}
\end{equation}
The two solutions for $\alpha $, corresponding to a faster and slower
falloff mode of the scalar for $r\rightarrow 0$, always exist for $d\geq 8$,
whereas for $d<8$ we must require $\theta\ge -d(d+1)/(8-d)$.

The general  solution to Eq. (\ref{h6a}) is given by a superposition of 
the slowest and fastest fall off modes:
\begin{equation}
\delta\phi= a(kr)^{\a_{2}¥}(1+{\cal O}(r^{2}¥))+
b (kr)^{\a_{1}¥}(1+{\cal O}(r^{2}¥)),
\label{g10}
\end{equation}
where $a,b$ are ${\cal O}(1)$ constants determined by the boundary 
conditions, we have taken the  $(t,x_{i})$-Fourier transform and
$k^{2}=-k_{0}^{2}+k_{i}k_{i}$.
The Green's function $G(k)$  for the scalar operator dual to the bulk  scalar 
field is given by the ratio of the coefficients of the $r^{\a_{1}¥}$ 
and $r^{\a_{2}¥}$ terms in Eq. (\ref{g10}) (see for instance 
\cite{Son:2002sd}), 
\begin{equation}
G(k)\sim k^{\alpha _{1}-\alpha _{2}\ },  \label{h10}
\end{equation}%
where $\alpha _{12}$ are given by Eq. (\ref{g9}). Taking the Fourier
transform, in the coordinate space we get the power-law form for the
two-point function for the scalar operator dual to a bulk scalar field with
exponential potential: 
\begin{equation}
\langle \mathcal{O}(x)\mathcal{O}(x^{\prime })\rangle =\frac{1}{|x-x^{\prime
}|^{d+1+\alpha _{1}-\alpha _{2}}}.  \label{l1a}
\end{equation}

It is also of interest to compute the two-point function (\ref{l1a}) for
small negative values of $\theta $: 
\begin{equation}
\langle \mathcal{O}(x)\mathcal{O}(x^{\prime })\rangle =\frac{1}{|x-x^{\prime
}|^{2(d+1)-(d-4)\theta/d}}.  \label{l1b}
\end{equation}

\section{Conclusions}

\label{sect:AF} In this paper we have analyzed  the thermodynamics
and the scaling symmetries of BB solutions of AdS Einstein-scalar
gravity in arbitrary dimensions for models with positive scalar squared mass
and a potential that has an  exponential asymptotic behavior. We have
generalized the results of Ref. \cite{Cadoni:2012uf}, which hold for two-dimensional
scalar branes, to branes of arbitrary spacetime dimensions.

We have been mainly concerned with an integrable model, which also arises as
compactification of black $p$-brane solutions of SUGRA theories. However,
the relevant features of this model can easily be extended to a broad class
of Einstein-scalar gravity models.

The striking features of these $d$-dimensional scalar BB solutions are an
unexpected phase diagram and  non-trivial behavior in the
ultraviolet regime of the holographically dual QFT, which is characterized by
hyperscaling violation. This generates an UV length scale which decouples in
the IR, where conformal invariance is restored. At high temperatures, 
when $\b^{2}<1/(d+1)$ or $\b^{2}>d+1$ the
scalar-dressed BB solution, with scale-covariant asymptotical behavior, 
becomes energetically preferred.

The hyperscaling violating phase is characterized by the two parameters
normally used for critical systems with hyperscaling violation, namely the
dynamical critical exponent $z$ and the hyperscaling violation parameter $%
\theta $.

The most important peculiarity  of our models is that for scalar 
black branes
that are stable at high temperatures, the hyperscaling parameter $\theta$ is 
always \textsl{negative}. In QFTs
with hyperscaling violation the scaling law for the free energy is that
pertinent to a CFT in $d-\theta$ dimensions. For positive $\theta$ we have
therefore a lowering of the effective dimensions.  This  is an 
important feature 
of the small temperature behavior of traditional
hyperscaling-violating critical systems \cite{Fisher:1986zz}. On the other
hand, the scalar BB brane solutions investigated in this paper  are 
characterized by a negative
hyperscaling-violation parameter $\theta$, producing 
a raising of the ``effective dimensions''.

It is important to notice in this context that the most general
compactification of $p$-brane solutions of SUGRA theories produces
hyperscaling violation in the dual QFT with both $\theta <0$ or $\theta >d$.
Both cases are consistent with the null energy condition for the bulk stress
energy tensor, but for $\theta >d$  the SAdS phase is always 
energetically preferred (see Sect. \ref{sect:AC}). On the other hand the
simplest diagonal ansatz (\ref{k8}) for the D-dimensional metric leads to BB
solutions with $\theta >d$.

We have also determined, for the case of negative $\theta $, the short
distance behavior of two-point functions for scalar operators of the QFT
dual to a bulk scalar field with an exponential potential. We have shown
that it has a power-law behavior. Our calculation completes the derivation
of Ref. (\cite{Dong:2012se}). In that paper the short distance, power-law, form of the
two-point functions for scalar operators dual to a scalar field with a mass
term potential  was determined only for positive $\theta $.

A puzzling point which still remains to be clarified is the 
holographic interpretation of the phase transition between the 
two bulk phases - the SAdS and the 
scalar brane phase.
The cross-over of the free energies for SAdS and scalar branes
observed in Sect. \ref{sect:AC}
seems to have a very different interpretation than
a conventional phase transition in the gravity/gauge theory correspondence, such as
for instance the Hawking-Page phase transition.

Usually, in the  gravity/gauge theory correspondence,  we fix the 
boundary conditions for the fields  and consider  two
distinct extensions into the bulk. The corresponding dual  solutions 
contribute to the same canonical ensemble of the QFT.
In the large-N limit the solution with  lower 
free energy is  energetically preferred.
On the other hand the two competing phases of the QFT holographically 
dual to the SAdS- scalar brane phases  seem to correspond to different
boundary QFTs. Therefore they do not  contribute to 
the same  canonical ensemble.

This is obviously  related to the unusual feature that the scalar
black brane solutions discussed in this  paper
exhibit hyperscaling violation in the UV and conformal
symmetry in the IR. In the conventional
setting where the solution has a UV fixed point and an emergent 
nonzero $\theta$ in the IR,  the holographic interpretation of 
the phase transition is not problematic. In this latter 
case the SAdS and 
the hyperscaling violating phase contribute to the same canonical ensemble. 

\begin{acknowledgments}
We thank P.\ Pani for useful discussions and illuminating comments. MS
gratefully acknowledges Sardinia Regional Government for the financial
support of his PhD scholarship (P.O.R. Sardegna F.S.E. Operational Program
of the Autonomous Region of Sardinia, European Social Fund 2007-2013 - Axis
IV Human Resources, Objective I.3, Line of Activity I.3.1).
\end{acknowledgments}

\bibliography{dw1}


\end{document}